\documentclass[12pt]{article}

\usepackage[vcentermath]{youngtab}
\usepackage{color}

\def\be{\begin{equation}}
\def\ee{\end{equation}}
\def\ba{\begin{eqnarray}}
\def\ea{\end{eqnarray}}

\usepackage[centertags]{amsmath}
\usepackage{amsfonts} 
\usepackage{amssymb}
\usepackage{amsthm}
\numberwithin{equation}{section} 
\def\ii{{\rm i}}
\newcommand{\ex}[1]{{\rm e}^{#1}} 

\begin{document}
\begin{titlepage}
\hfill \hbox{CERN-PH-TH/2013-244}
\vskip 0.1cm
\hfill \hbox{NORDITA-2013-88}
\vskip 0.1cm
\hfill \hbox{QMUL-PH-13-11}
\vskip 1.5cm
\begin{flushright}
\end{flushright}
\vskip 1.0cm
\begin{center}
{\Large \bf  The leading eikonal operator  in string-brane  \\
 scattering at high energy 
 }
 
  \vskip 1.0cm {\large Giuseppe
D'Appollonio$^{a }$, Paolo Di Vecchia$^{b, c}$,
Rodolfo Russo$^{d, e}$, \\
Gabriele Veneziano$^{f, g}$ } \\[0.7cm]
{\it $^a$ Dipartimento di Fisica, Universit\`a di Cagliari and
INFN\\ Cittadella
Universitaria, 09042 Monserrato, Italy}\\
{\it $^b$ The Niels Bohr Institute, University of Copenhagen, Blegdamsvej 17, \\
DK-2100 Copenhagen, Denmark}\\
{\it $^c$ Nordita, KTH Royal Institute of Technology and Stockholm University, \\Roslagstullsbacken 23, SE-10691 Stockholm, Sweden}\\
{\it $^d$ Queen Mary University of London, Mile End Road, E1 4NS London, United Kingdom}\\
{\it $^e$ Laboratoire de Physique Th\'eorique de L'Ecole Normale Sup\'erieure\\
24 rue Lhomond, 75231 Paris cedex, France}\\
{\it $^f$ Coll\`ege de France, 11 place M. Berthelot, 75005 Paris, France}\\
{\it $^g$Theory Division, CERN, CH-1211 Geneva 23, Switzerland}
\end{center}

\begin{abstract}
In this paper we present two (a priori  independent) derivations of the eikonal operator in string-brane scattering. The first one is  obtained by summing   surfaces with any number of boundaries, while in the second one the eikonal operator  is derived  from the   three-string vertex in a suitable  light-cone gauge.   This second derivation shows  that the bosonic oscillators present in the leading eikonal operator  are to   be identified with the string bosonic oscillators in a suitable  light-cone gauge, while the first one shows that it exponentiates recovering unitarity. 
This paper  is a review of results obtained in Refs.~\cite{ddrv2} and \cite{D'Appollonio:2010ae}. 
\end{abstract}

\vspace{.5cm}

\begin{center}
Talk given  by one of us (PDV) at  the conference BUDS2013, Breaking of supersymmetry and Ultraviolet Divergences in extended Supergravity, National Laboratories of Frascati,  
 March 25-28, 2013. 
\end{center}
\vspace{1cm}

\end{titlepage}


\section{Introduction}
\label{intro}

High energy scattering in the Regge limit in superstring theory has been investigated since more than 25 years. It was originally studied in elastic string-string collisions~\footnote{For a complete list of references see Ref.~\cite{ddrv2}.} and has more recently been extended to the elastic scattering of a closed string on a $Dp$-brane~\cite{D'Appollonio:2010ae}. Due to the fact that, in the Regge limit, the amplitude is dominated by the exchange of the leading Regge trajectory that has  the graviton as the lowest state, one gets a lowest order (sphere or disc) amplitude that diverges
with the energy violating unitarity at high energy. Unitarity is restored by adding higher order corrections (torus or annulus etc.) and summing  them up. In this way, while in field theory one gets an exponential with a phase divergent at high energy that is consistent with unitarity, 
what  one obtains in string theory can be written in terms of  an infinite set of bosonic  oscillators, introduced to write the amplitude in a simple and compact form,  
and    is called the leading eikonal operator. 

This construction poses, however,  various problems. 
What are these bosonic oscillators? Are they connected to the bosonic oscillators of superstring theory?
Since we are studying superstring theory, why don't we get also fermionic oscillators?        
Although the connection of these oscillators with the string  oscillators was unclear, it was believed that they were somehow directly related to the string bosonic oscillators. Evidence of this connection came from a paper by Black and Monni~\cite{Black:2011ep} where the disk amplitude for the production of  massive  states,  lying on the leading Regge trajectory, from the scattering of a massless state on a $Dp$-brane was computed and found to agree with what one gets from the eikonal operator.  It turns out, however,  that this comparison is more subtle because one has to take into account that the longitudinal polarization of the massive state gets enhanced at high energy  pretty much as the longitudinal component of the gauge boson $W^{\pm}$ in the Standard Model without  the Higgs boson.   

In a recent paper~\cite{ddrv2}  the problems  raised above were clarified showing that the bosonic oscillators appearing in the eikonal operator are the bosonic oscillators of superstring in a suitable 
light-cone gauge and that  the fermionic oscillators are not relevant at high energy. 
Furthermore, it was shown how to correctly treat the longitudinal polarization of the massive state.
This  means that, if we scatter a massless state on a $Dp$-brane, we  
produce, at high energy, only  massive states involving  an arbitrary number of bosonic oscillators together with only  the fermionic oscillators    already  present in the  massless state.   
Actually, the analysis of Ref.~\cite{ddrv2} is more general because it provides the production 
amplitude in the Regge high energy limit of an arbitrary state of superstring theory from the scattering of an  arbitrary state on the $Dp$-brane.
In particular,  it has been shown~\cite{ddrv2}  that the leading eikonal operator can be directly derived starting 
from the three-string light-cone vertex (either in the form of Green-Schwarz or in that of Ramond-Neveu-Schwarz) and then  inserting in  one of the three legs  the string propagator and by closing it
with the boundary state that takes care of  the presence of the $Dp$-branes. This provides a direct
construction of the leading eikonal operator from the string operator formalism.  The aim of this talk is to present these recent results. In Ref.~\cite{ddrv2} the leading eikonal operator has been also constructed by using a covariant formalism in terms of the Reggeon  vertex operator, but this will not be reviewed here. 

The content of this paper is the following. In Sect. \ref{eikonal,I} we derive the eikonal operator as it was originally constructed in Ref.~\cite{D'Appollonio:2010ae} starting from the scattering amplitudes.  
In Sect. \ref{1massivelevel}  we give a description of the physical spectrum of the first massive level  in the two light-cone formalisms (GS and RNS) and in the covariant formalism.  Then, interpreting the bosonic oscillators of the eikonal operator as the light-cone bosonic oscillators of string theory,  we show that, 
at high energy, the  states that can be produced by the scattering of a graviton on a $Dp$-brane, are only those  of the type $A_{-1;j} |i, 0\rangle$, while those of the type $Q_{-1; a} | \dot{a}, 0\rangle$ are not. This is consistent with what one gets from the eikonal operator that does not contain any fermionic oscillator. In Sect. \ref{eikonal,II}
we show how to derive the eikonal operator from the light-cone three-string vertex and the boundary state. Finally, an Appendix with a discussion of the kinematics of the scattering process
is added at the end of the paper.

\section{The eikonal operator  I}
\label{eikonal,I}

In this section we derive  the leading eikonal operator from the elastic scattering of a massless state 
of superstring theory  on a Dp-brane, following   Ref.~\cite{D'Appollonio:2010ae}. 
The starting point is the disk  amplitude given by:
\begin{eqnarray}
&&{\cal{A}}_{1} (E, t) \sim\langle 0 | \int \frac{d^2 z_1 d^2 z_2}{dV_{abc}} W_1  (z_1 , {\bar{z}}_1 ) 
W_2  (z_2 , {\bar{z}}_2 )  | B \rangle \nonumber \\
&& = - \frac{ \pi^{\frac{9-p}{2}} R_{p}^{7-p}}{\Gamma ( \frac{7-p}{2} )} \,\,
 {\cal{K}} (p_1 , \epsilon_1 ; p_2 , \epsilon_2 ) \frac{\Gamma (- \alpha' E^2) 
\Gamma ( - \frac{\alpha'}{4}t)}{\Gamma (1-
\alpha' E^2 -\frac{\alpha'}{4}t )}
\label{disk}
\end{eqnarray}
where 
\begin{eqnarray}
R_p^{7-p} =  g N \frac{(2 \pi \sqrt{\alpha '})^{7-p}}{(7-p) V_{S^{8-p}}} \ , 
\hspace{1cm} V_{S^n} = \frac{2 \pi^{\frac{n+1}{2}}}{\Gamma(\frac{n+1}{2})} \,\, ,
\label{Rp}
\end{eqnarray}
$W_1$ and $W_2$ are the vertex operators of a massless state and   $|B\rangle$ is the boundary state that identifies the right with the left oscillators and imposes  Dirichlet (Neumann) boundary conditions along  the directions transverse (longitudinal)  to the world-volume of the stack of $N$ parallel $Dp$-branes. 
The scattering is described by two Mandelstam-like  variables:
\begin{eqnarray}
t =  - (p_{1\perp} + p_{2\perp})^{2}  = - 4E^2 \sin^2 \frac{\Theta}{2}
~~;~~~s=  E^2 =  | p_{1\perp }|^{2} =  | p_{2\perp}|^{2}
\label{ts}
\end{eqnarray}
$\Theta$  is the angle between the  $(9-p)$-dim  vectors ${p_{1\perp}}$ and $-{p_{2\perp}}$.

Along the directions of the world-volume of the $Dp$-branes, there is conservation of energy and momentum: 
\begin{eqnarray}
(p_1 + p_2)_{\parallel} =0~~~~;~~~~~p_{1}^{2} = p_{2}^{2} =0
\label{longcons}
\end{eqnarray}
The amplitude has simultaneously poles for  $E^2$  such that $1 + \alpha' E^2 =n \,(n=1,2 \dots)$   corresponding to  open  strings exchanged in the $s$-channel  and poles for  $t$ such that 
$2 + \frac{\alpha'}{2}t =2m (m=1,2 \dots)$ corresponding to closed strings exchanged in the $t$-channel. At high energy:
\begin{eqnarray}
 {\cal{K}} (p_1 , \epsilon_1 ; p_2 , \epsilon_2 ) = (\alpha' E^2)^2 Tr (\epsilon_1 \epsilon_{2}^{t} )
\label{he}
\end{eqnarray}
and the amplitude has  Regge behaviour for $ \alpha ' s >> \alpha' t \sim 0$ ($s \equiv E^2$):
\begin{eqnarray}
T_1 (E,t) \equiv \frac{ {\cal{A}}_1 (E,t)}{2E}  =  
\frac{  R^{7-p}_{p} \pi^{\frac{9-p}{2}} }{\Gamma ( \frac{7-p}{2} ) } \,\,
\frac{ \pi {\rm e}^{- i \frac{\alpha'}{4}t} { ( {\alpha' } s  )^{1 + \frac{\alpha'}{4}t }}}{2E \,\,
\sin (\pi \frac{\alpha'}{4} (-t) )\,\, \Gamma ( 1+ \frac{\alpha't}{4})} 
\label{T1}
\end{eqnarray}
$T_1$  has  a real  and an imaginary part. The real part describes the scattering of the closed string on the $Dp$-brane, while the imaginary part  describes the absorption of the closed string by the $Dp$-brane. When $\alpha' \rightarrow 0$ the real part  reduces to the field theoretical result
(graviton exchange), while for $\alpha' \neq 0$ we have the graviton exchange dressed with string corrections. Notice that the imaginary  part is a pure string correction that, however, is not relevant at very large impact parameter because it is not divergent at $t=0$ as the real part. The disk amplitude in 
Eq. (\ref{T1}) diverges at high energy and violates unitarity. In order to restore unitarity we have to
include higher order corrections and sum them up. Before we proceed further it is instructive to  write the corresponding amplitude that one gets  in the bosonic string for  the elastic scattering of a closed string tachyon on a $Dp$-brane:
\begin{eqnarray}
{\cal{A}}_1 \sim \frac{\Gamma ( -1 - \alpha' s) \Gamma ( - \frac{\alpha't}{4} -1)}{\Gamma ( - \alpha' s - 
 \frac{\alpha't}{4} -2)}= \frac{ \Gamma ( - \alpha_{open} (s)) \Gamma ( - \frac{ \alpha_{closed} (t)}{2}  )}{
 \Gamma ( - \alpha_{open} (s)  - \frac{ \alpha_{closed} (t)}{2}  )}
\label{A1bos}
\end{eqnarray}
where $\alpha_{open} (s) = 1 + \alpha' s$ and $\alpha_{closed} (t) = 2 + \frac{\alpha'}{2}t$. 
It has the same form as the original Veneziano model except having two different trajectories in the two channels: one corresponding to the open string and the other to the closed string. 

The next diagram is the annulus diagram that  is given by:
\begin{eqnarray}
{\cal{A}}_2  = {\cal{N}} \int d^2 z_1 d^2 z_2 
\sum_{\alpha, \beta} {}_{\alpha \beta} 
\langle B | W_{1}^{(0)} (z_1 , {\bar{z}}_1)  W_{2}^{(0)} (z_2 , {\bar{z}}_2) D
| B \rangle_{\alpha, \beta} 
\label{Am}
\end{eqnarray}
${\cal{N}}$ is a
 normalization factor and $\sum_{\alpha, \beta}$ is the sum over the spin structures. 

The sum over the spin structures can be explicitly performed obtaining in practice only 
the contribution of the bosonic degrees of freedom without  the  bosonic partition function.

The final result is rather explicit. In the closed string channel the coefficient of the term with 
${\rm Tr} ( \epsilon_1 \epsilon_{2}^{T})$ 
(relevant at high energy) of   the annulus  is equal to:
\begin{eqnarray}
{\cal A}_2(s, t) &=& \frac{\pi^3 (\alpha'  s)^2 }{\Gamma^2\left (
\frac{7-p}{2} \right )} \frac{R_{p}^{14-2p}}{(2
\alpha')^{\frac{7-p}{2}}} \nonumber \\
&\times&
\left[2 \int_{0}^{\infty} \frac{d \lambda }{\lambda^{ \frac{5-p}{2}}} \,
\int_{0}^{\frac{1}{2}} d \rho_1 \int_{0}^{\frac{1}{2}} d \rho_2
\int_{0}^{1} d \omega_1 \int_{0}^{1} d \omega_2 \
{\cal I}\right]  
\label{anulus}
\end{eqnarray}
where
\begin{eqnarray}
{\cal{I}} \equiv {\rm e}^{- \alpha' s V_s - \frac{\alpha'}{4} t V_t }~~~;~~~z_{1,2} \equiv {\rm e}^{2 \pi  
(- \lambda \rho_{1,2} +
i \omega_{1,2})}
\label{finalissimoclo}
\end{eqnarray}
and 
\begin{eqnarray}
V_s  =-  {2 \pi} \lambda  \rho^{2} + \log \frac{\Theta_1 ( i \lambda (\zeta+ \rho ) | i \lambda) \Theta_1 ( i \lambda( \zeta- \rho|) i \lambda)  }{ \Theta_1 (i\lambda  \zeta +  \omega) | i \lambda) 
\Theta_1 (i \lambda  \zeta-  \omega)| i \lambda) }
\label{Vsabla}
\end{eqnarray}
 and
\begin{eqnarray}
V_t  =8 \pi \lambda \rho_1 \rho_2 + \log \frac{\Theta_1 (i \lambda  \rho + \omega) | i \lambda) \Theta_1 (i\lambda  \rho - \omega)| i \lambda)  }{ \Theta_1 (i \lambda  \zeta + \omega) | i \lambda) 
\Theta_1 (i \lambda  \zeta -  \omega)| i \lambda) }
\label{Vtabla}
\end{eqnarray}
with $\rho  \equiv \rho_1 - \rho_2 ~~;~~\zeta = \rho_1 + \rho_2 ~~;~~
\omega  \equiv \omega_1 - \omega_2$.

The high energy behaviour ($E  \rightarrow \infty$) of the annulus diagram  can be studied, by the saddle point technique, looking for points where $V_s$  vanishes.
This happens for $\lambda \rightarrow \infty$ and $\rho  \rightarrow 0$.

Performing the calculation one gets the leading term for $E \rightarrow \infty$:
\begin{eqnarray}
&&\frac{{\cal{A}}_{2}^{(3)} (E, t)}{2E}  \rightarrow \frac{i}{2}  \prod_{i=1}^{2} \left[ \int \frac{d^{8-p} {\bf k_i} }{(2\pi)^{8-p}} 
\frac{{\cal{A}}_{1} (E, t_i )}{2E}  \right]  \nonumber \\
&& \times
\delta^{(8-p)} ( \sum_{i=1}^{2} k_i - q)
\,\, V_2  (t_1 , t_2 ,t)~~~;~~t_i \equiv - {\bf k}^2_{i}~~;~~ t= - {\bf q}^2
\label{A2}
\end{eqnarray}
where
\begin{eqnarray}
V_2  (t_1 , t_2 ,t) = \frac{\Gamma ( 1 + \frac{\alpha'}{2} \left( t_1 + t_2 -t \right) ) }{\Gamma^2 
( 1 + \frac{\alpha'}{4} \left( t_1 + t_2 -t \right) )} 
\label{V2}
\end{eqnarray}
In order to find the complete leading eikonal operator we  write it 
in a more suggestive way,  in terms of an infinite set of  $(8-p)$-dim bosonic 
oscillators:
\begin{eqnarray}
V_2  (t_1 , t_2 ,t)  = \langle 0| \prod_{i=1}^{2}\left[  \int_{0}^{2 \pi} \frac{d\sigma_i}{2 \pi}  
:{\rm e}^{ i {\mathbf k}_i \cdot X( \sigma_i )}: \right] |0 \rangle
\label{V22}
\end{eqnarray}
where
\begin{equation}
\label{Xex}
\hat{X}(\sigma) = i \sqrt{\frac{\alpha'}{2}} \sum_{n\not=0}
\left(\frac{\alpha_n}{n} e^{i n \sigma}+
\frac{\tilde{\alpha}_n}{n} e^{-i n \sigma}\right) 
\end{equation}
The two vacuum states correspond to the two external massless states (states with no bosonic excitations: $( \epsilon_{\mu \nu}  \psi_{-\frac{1}{2} }^{\mu}  {\tilde{\psi}}_{-\frac{1}{2} }^{\nu} |0 \rangle$).

Then the leading order from the annulus can be written as follows:
\begin{eqnarray}
&&\frac{{\cal{A}}_{2}^{(3)} (E, t)}{2E}  \rightarrow \frac{i}{2}\prod_{i=1}^{{2}} \left[ 
\int \frac{d^{8-p} {\bf k}_i }{(2\pi)^{8-p}} 
\frac{{\cal{A}}_{1} (E, - {\bf k}_{i}^{2} )}{2E} \right]  \delta^{(8-p)} ( \sum_{i=1}^{{2}} {\bf k}_i - {\bf q})  
\nonumber \\
&& \times
\langle 0| \prod_{i=1}^{{2}} \left[  \int_{0}^{2 \pi} \frac{d\sigma_i}{2 \pi} 
:{\rm e}^{ i {\mathbf k}_i \cdot X( \sigma_i )}: \right] |0 \rangle
\label{leadannu}
\end{eqnarray}
where the two vertex operators correspond to the two leading Reggeons
exchanged in the two $t$-channels: $t_1$ and $t_2$.

It can be naturally generalized to the leading term coming from a 
surface with  $h$ boundaries:
\begin{eqnarray} 
\label{ah38}
&&\frac{ {\cal A}_{h}^{(h+1)} (s, t)}{2E} \sim \frac{i^{h-1} }{h!} 
\prod_{i=1}^{{h} } \left[ \int \frac{d^{8-p} {\bf
k}_i}{(2 \pi)^{8-p}} \ \frac{ {\cal A}_1(s,- {\bf k}_{i}^{2})}{2E} \right]   
\nonumber \\
&& \times \delta^{(8-p)} ( \sum_{i=1}^{{h}} {\bf k}_i - {\bf q})
\,\,  \langle 0| \prod_{i=1}^{{h} } \left[
\int_{0}^{2 \pi} \frac{d\sigma_i}{2 \pi} 
:{\rm e}^{ i {\mathbf k}_i \cdot X( \sigma_i )}:   \right]|0 \rangle
\label{hbou}
\end{eqnarray}
Going to impact parameter space
\begin{eqnarray}
&&
i \frac{{\cal A}^{(h+1)}_h(s, {\bf b}) }{2E} =
\int  \frac{d^{8-p} {\bf q}}{(2 \pi)^{8-p}} \ e^{i {\bf b} {\bf q}}
\  i  \frac{{\cal A}^{(h+1)}_h(s, t)  }{2E} \nonumber \\
&& = \frac{i^h}{h!}  
\prod_{i=1}^{{h}} \left[  \int \frac{d^{8-p} {\bf k}_i}{(2\pi)^{8-p}}
\frac{{\cal{A}}_1 (s, - {\bf k}_{i}^{2})}{2E} \right]   
\nonumber \\
&& \langle  0 | 
  \prod_{i=1}^{{h}} \left[ \int\limits_{0}^{2 \pi} \frac{d \sigma_i }{2 \pi} 
: {\rm e}^{ i {\bf k}_i ({\bf b} + \hat{X} (\sigma_i ) ) }:
\right] |0 \rangle 
\label{ah40}
\end{eqnarray}
and summing all contributions:
\begin{eqnarray}
\sum_{h=1}^\infty \frac{{\cal A}^{(h+1)}_h(s, {\bf b})}{2E} \sim
\langle 0 |\frac{1}{i} \left [ e^{2 i \hat{\delta}(s, b)}
- 1 \right ] | 0 \rangle 
\ , 
\label{eo}
\end{eqnarray}
we get the leading  eikonal operator
\begin{eqnarray}
2 \hat{\delta}(s, b) &=&
\int\limits_0^{2\pi} \frac{d \sigma}{2 \pi}
\int \frac{d^{8-p} {\bf k}}{(2
\pi)^{8-p}} \, \frac{{\cal A}_{1} (s,-{\bf k}^2)}{2 E}
: e^{i {\bf k} ( {\bf b} + \hat{{\bf X}}(\sigma))} : \nonumber \\
&=& \int\limits_0^{2\pi} \frac{d \sigma}{2 \pi}
\frac{: {\cal A}_{1}
\left (s, {\bf b} + \hat{{\bf X}}(\sigma) \right ):
}{2E}    
\label{ep}
\end{eqnarray}
The final result that includes all string corrections
is obtained from the field theoretical one with the substitution:
\begin{eqnarray}
{\bf b}  \Longrightarrow {\bf b} +  {\hat{\bf X}}~~;~~\hat{\bf{X}}(\sigma) = i \sqrt{\frac{\alpha'}{2}} \sum_{n\not=0}
\left(\frac{\alpha_n}{n} e^{i n \sigma}+
\frac{\tilde{\alpha}_n}{n} e^{-i n \sigma}\right) 
\label{substi}
\end{eqnarray}
and normal ordering.

This is the way that  the leading eikonal  operator was originally constructed both in string-string and string-brane scattering. From this derivation it is not clear what the bosonic oscillators represent. It was, however,  somehow believed that, when the eikonal operator is saturated with a couple of physical states, it will reproduce the high energy behaviour of their scattering amplitude. 

For the states of  the leading Regge trajectory it has been shown~\cite{Black:2011ep}  that the quantity
\begin{eqnarray}
\int \frac{d^{8-p} {\bf k}  }{(2\pi)^{8-p}}  \frac{ {\cal{A}}_{1} (E, - {\bf k}^{2} )}{2E} 
 \delta^{(8-p)} (  {\bf k} - {\bf q}) \,\,\,
\langle 0| \int_{0}^{2 \pi} \frac{d\sigma}{2 \pi} 
:{\rm e}^{ i {\mathbf k} \cdot X( \sigma )}:   | \lambda \rangle
\label{mass1}
\end{eqnarray}
reproduces the high energy behaviour of the  disk amplitude involving a massless state 
($\langle 0 |$) and a state of the leading Regge trajectory ($| \lambda \rangle$). It turned out, however, that this computation is more subtle because the longitudinal polarization of the massive state gets enhanced at high energy.  The annulus diagram for a massless state and  an excited state of the leading Regge trajectory has also been  computed~\cite{Bianchi:2011se}. 

In any case, the problem of the nature of the bosonic oscillators present in the eikonal operator remains. Given the fact that in string-string collisions they are along the eight directions orthogonal to both the time and  the direction of the fast moving string and similarly in string-brane collisions they are along the $8-p$ transverse directions again orthogonal to the time and  to the direction of the fast moving string, strongly suggests that they should be interpreted as the string bosonic oscillators in the light-cone gauge. But even so, why does the eikonal operator  not contain the fermionic oscillators?

Putting this problem for a moment aside,  in the next section we compute the amplitude for the production of a massive state belonging to the first excited level of superstring theory from the scattering of a graviton on a $Dp$-brane and we compare with what one gets from the eikonal operator. We will show that, in agreement with the eikonal operator, we
produce, at high energy,  only   excited states of the graviton ($|i\rangle |\tilde{i} \rangle$)  of the type $ A_{-1, j}|i\rangle  {\tilde{A}}_{-1;{\tilde{j}}}|\tilde{i} \rangle$.  The remaining massive states
of the type $ Q_{-1, b}|a\rangle  {\tilde{Q}}_{-1;{\tilde{b}}}|{\tilde{a}} \rangle$,  
$A_{-1, j}|i\rangle   {\tilde{Q}}_{-1;{\tilde{b}}}|{\tilde{a}} \rangle$  and $Q_{-1, b}|a\rangle 
 {\tilde{A}}_{-1;{\tilde{j}}}|\tilde{i} \rangle$
are not produced at high energy.

\section{States of the first massive level produced  at high energy}
\label{1massivelevel}

In order to understand the problems listed at the end of the last section, in this section we 
consider the production of a massive state, belonging to the first massive level, from the scattering of  a massless state on a $Dp$-brane and we study which of the $128 \times 128$ bosonic states  are produced at high energy
in the Regge limit.  This section is divided in three  subsections. In the first one we compare  the spectrum of physical states at the first excited level in the Green-Schwarz light-cone formalism,
in the RNS light-cone formalism and in the covariant formalism. We introduce also the DDF operators that connect the states in the light-cone RNS with those in the covariant formalism.
In the second short subsection we compute the three-point amplitudes involving two gravitons and 
a  bosonic state of the first excited level. Finally,
in the third  subsection, we compute the inelastic amplitude for the production of the states
of the first excited level and we check which of them are produced at high energy.

\subsection{Spectrum of the first excited level}
\label{spectrum}

In this subsection we discuss the spectrum of physical  states of the first massive level in closed superstring theory in  the two light-cone gauges (Green-Schwarz (GS) and Ramond-Neveu-Schwarz (RNS)) and in the covariant formalism. Any closed string state is  a product of a state with left   moving oscillators times a state with right moving oscillators. In the following we discuss only the states
with one type of oscillators.  Those with the other type of oscillators can be obtained exactly in the same way. 

 \begin{enumerate}

\item {\bf GS light-cone}

In the GS light-cone the bosonic physical states at the first massive level are the following:
\begin{eqnarray}
&& \alpha_{-1}^{i} | j \rangle  \Longrightarrow   64 \,\,\, states \nonumber \\
&& Q_{-1}^{a} | b\rangle      \Longrightarrow   64 \,\,\, states
\label{GS}
\end{eqnarray}
where $i,j=1 \dots 8$ are vector indices and $a,b=1 \dots 8$ are spinor indices of $SO(8)$. 

\item {\bf RNS light-cone}

In the RNS light-cone the bosonic physical states are the following:
\begin{eqnarray}
&& A_{-1}^i B_{-\frac{1}{2}}^j |0\rangle   \Longrightarrow   64 \,\,\, states \nonumber \\
&& B_{-\frac{3}{2}}^i |0\rangle   \Longrightarrow   8 \,\,\, states \nonumber \\
&&B_{-\frac{1}{2}}^i B_{-\frac{1}{2}}^j B_{-\frac{1}{2}}^k
 |0\rangle   \Longrightarrow   56 \,\,\, states 
\label{RNS}
\end{eqnarray}
where $i,j,k=1 \dots 8$ are vector indices of $SO(8)$.  The states in the first line of Eq. (\ref{GS}) correspond to those in the first line of Eq. (\ref{RNS}), while the states in the second line of Eq. (\ref{GS}) correspond to those in the second and third line of Eq. (\ref{RNS}).

\item {\bf Covariant formalism}

In the covariant formalism the physical states in the center of mass frame \\ ($p=(M,\vec{0})$)  are:
\begin{eqnarray}
&& T^{IJ} =  \left( \alpha_{-1}^{I} \psi_{-\frac{1}{2}}^{J} +  \alpha_{-1}^{J} \psi_{-\frac{1}{2}}^{I}
- \frac{2}{9} \eta^{IJ} \eta^{HK}  \alpha_{-1}^{H} \psi_{-\frac{1}{2}}^{K} \right) |0, p  \rangle  
 \Longrightarrow   44  \,\,\, states \nonumber \\
&& V^{IJK} =  \psi_{-\frac{1}{2}}^{I}  \psi_{-\frac{1}{2}}^{J}  \psi_{-\frac{1}{2}}^{K} | 0, p \rangle 
  \Longrightarrow   84  \,\,\, states 
\label{COVfo}
\end{eqnarray}
where $I,J,K,H = 1 \dots 9$ are vector indices of $SO(9)$. We can decompose the 9-dim indices $I=i,v;J=j,v$ in 8-dim indices and a longitudinal one that we call $v$:
\begin{eqnarray}
T^{ij}      \Longrightarrow   36  \,\,\, states~~~~;~~~~~T^{iv}   \Longrightarrow   8  \,\,\, states \nonumber \\
V^{ijk}   \Longrightarrow   56  \,\,\, states ~~~~~;~~~~~~
V^{ijv}  \Longrightarrow   28  \,\,\, states 
\label{decompo11}
\end{eqnarray}
$T^{ij}$ and $V^{ijv}$ correspond to the  64 states in the first line of Eq. (\ref{RNS}), while the others correspond to those in the second and third line of Eq. (\ref{RNS}). The two states in Eq.
(\ref{COVfo}) can be given a covariant  $SO(1,9)$ form by a boost, In this way one gets the following states: 
\begin{eqnarray}
|\phi_1 \rangle = 
T^{\alpha \rho}_{ \alpha'  \rho'} \alpha_{-1}^{\rho'}  \psi_{-\frac{1}{2}}^{\alpha'}  |0, p  \rangle
\label{Tar}
\end{eqnarray}
where  
\begin{eqnarray}
T^{\alpha \rho}_{ \alpha'  \rho'}=  ( \eta_{\perp})^{\rho}_{ \rho'} ( \eta_{\perp})^{\alpha}_{ \alpha'} +  (\eta_{\perp})^{\rho}_{ \alpha'}  (\eta_{\perp})^{\alpha}_{ \rho'} - \frac{2}{9}  
\eta_{\perp}^{\rho \alpha}  \eta_{\perp \alpha'  \rho'}  ~;~
(\eta^{\mu \nu}_{\perp} = \eta^{\mu \nu} - \frac{p^\mu p^\nu}{p^2})
\label{mas11}
\end{eqnarray}
and 
\begin{eqnarray}
|\phi_2 \rangle =
\eta^{\rho}_{\perp \rho'} \eta^{\sigma}_{\perp \sigma'} \eta^{\tau}_{\perp \tau'} \psi_{-\frac{1}{2}}^{\rho'}  \psi_{-\frac{1}{2}}^{\sigma'} \psi_{-\frac{1}{2}}^{\tau'}|0, p \rangle  
\label{3psi}
\end{eqnarray}
It can be shown  that the two  states in Eqs. (\ref{Tar}) and (\ref{3psi}) are physical states:
\begin{eqnarray}
G_{\frac{1}{2}}  | \phi_{1,2} \rangle =G_{\frac{3}{2}}  | \phi_{1,2} \rangle =0 
\label{GG1231}
\end{eqnarray}
\end{enumerate}
The connection between the RNS oscillators in the light-cone gauge and those in the covariant formalism is provided by the DDF operators~\cite{DelGiudice:1971fp}. In the case of superstring they can be found in Ref.~\cite{Hornfeck:1987wt} and they are reviewed in Ref.~\cite{ddrv2}. In particular, as discussed in Ref.~\cite{ddrv2}, one gets for the states at the first massive level made with one A and one B oscillators:
\begin{eqnarray}
A_{-1,j} B_{-\frac{1}{2},k} |p_T  , 0\rangle &=& \left\{
\frac{1}{2} \left[  \alpha_{-1}^{j}
    \psi_{-\frac{1}{2}}^{k} +  \alpha_{-1}^{k} \psi_{-\frac{1}{2}}^{j} 
    - \frac{\delta^{jk}}{3} \left(\sum_{i=1}^{8}  \alpha_{-1}^{i}
      \psi_{-\frac{1}{2}}^i - 2 \alpha_{-1}^v \psi_{-\frac{1}{2}}^v
    \right)\right]   \right.  \nonumber \\
&+& \left.   \frac{1}{\sqrt{2}}  
(v \psi_{-\frac{1}{2}}) \psi_{-\frac{1}{2}}^{j} \psi_{-\frac{1}{2}}^{k}   \right\} |p ;0 \rangle~~;~~j,k =1 \dots  8~.
\label{AB}
\end{eqnarray}
where $(\epsilon_{j})_{\mu} \psi_{-\frac{1}{2}}^\mu \equiv  \psi_{-\frac{1}{2}}^j$, $(\epsilon_{j})_{\mu} 
\alpha_{-1}^{\mu}$, $p_T$ is the momentum of the tachyon present in the DDF state
and $v$ is the longitudinal polarization of the massive state that is orthogonal to the momentum $p$.  Analogously, one can also compute the connection with the covariant states of the other  two DDF states: $B_{-\frac{1}{2},i}  B_{-\frac{1}{2},j} B_{-\frac{1}{2},k}
|0,p_T \rangle$ and $B_{-\frac{3}{2},i} |0, p_T \rangle$.   

\subsection{Three-point amplitudes}
\label{3point}

In this subsection we provide  the three-point amplitude, in the covariant formalism, involving two gravitons  and  one of the  states of the first massive level.  In a closed string theory the amplitude is the product  of two amplitudes of open string theory, one for the left movers and the other for the right movers. Here, we quote only the result for the left movers.

For the massive state  in Eq. (\ref{mas11}) one gets: 
\begin{eqnarray}
A^{\mu ; IJ}_{\nu} (\phi_1) \sim    \epsilon_{\alpha \rho}^{IJ}  \frac{\alpha'}{2}   
\left[  \eta^{\mu \alpha } p_{3}^{\rho} p_{1}^{\nu}  - 
\eta^{\nu \alpha} p_3^{\rho}  p_{3}^{\mu} + \eta^{\mu \nu} p_{3}^{\alpha} p_3^{\rho}+ 
  \eta^{\mu \alpha} \eta^{\nu \rho}  \right]  
\label{V001covz}
\end{eqnarray}
where $p_1$ and $p_3$ are the momenta of the two gravitons and we have assumed that the polarization matrix  is symmetric, traceless and orthogonal to the four-momentum $p_2$ of the massive state:
\begin{eqnarray}
p_2^{\alpha}  \epsilon_{\alpha \rho}^{IJ} = \eta^{\alpha \rho} \epsilon_{\alpha \rho}^{IJ}=0
\label{consetrac}
\end{eqnarray}
For  the state in Eq. (\ref{3psi}) one gets:
\begin{eqnarray}
&&A^{\mu ; I,J,K}_{\nu} (\phi_2 ) \sim
\epsilon_{\rho \sigma \tau}^{ IJK}  \sqrt{ \frac{\alpha'}{2}}  \left[ \eta^{\nu \rho} \left( p_{3}^{\sigma} \eta^{\mu \tau} -
p_{3}^{\tau} \eta^{\mu \sigma}  \right)  \right. \nonumber \\  
&& \left. 
- p_{3}^{\rho} \left(  \eta^{\nu \sigma}  \eta^{\mu \tau}  - \eta^{\mu \sigma} \eta^{\nu \tau}      \right) + \eta^{\mu \rho} \left( \eta^{\nu \sigma}
p_{3}^{\tau} - \eta^{\nu \tau} k_{3}^{\sigma}   \right) \right]
\label{3psiam}
\end{eqnarray}
In this case the polarization is completely antisymmetric and orthogonal to the four-momentum of the massive state $p_2$. 
The indices $\mu$ and $\nu$ must be saturated with the  left moving part of the polarization of the two gravitons. We have assumed that all three states are incoming: $p_1+p_2+p_3 =0$.

\subsection{Inelastic amplitudes}
\label{inelastic}

In this subsection we use the three-point amplitudes of the previous section to compute the inelastic amplitude where the  graviton with momentum $p_1$ scatters on a $Dp$-brane producing a massive state with momentum $p_2$. This can be done by considering  the product of any of the two amplitudes (one for the right movers and the other for the left movers) constructed above and by saturating  the indices $\nu$ and ${\bar{\nu}}$ of the  graviton with momentum $p_3$  first  with the graviton propagator  in the De Donder gauge:
\begin{eqnarray}
D^{\nu \lambda; {\bar{\nu}}{\bar{\lambda}}} = \frac{\eta^{\nu \lambda} \eta^{{\bar{\nu}} {\bar{\lambda}}}  +   \eta^{\nu {\bar{\lambda}}} \eta^{{\bar{\nu}} {{\lambda}}}  - \frac{1}{4} \eta^{\nu {\bar{\nu}}} \eta^{\lambda {\bar{\lambda}}}   }{2 p_3^2 }
\label{gravipropa}
\end{eqnarray}
and then   with the coupling of the graviton to the $Dp$-brane given by 
\begin{eqnarray}
 \frac{1}{2}T_p \frac{ \eta^{\lambda {\bar{\lambda}}}+ R^{\lambda {\bar{\lambda}}}}{2}~~;~~T_p = \sqrt{\pi} (2\pi \sqrt{\alpha'})^{3-p}
\label{couplingbra}
\end{eqnarray}
where $R$ is the reflection matrix:
\begin{equation}
\label{reflema}
R^\mu_{~\nu}  =  \delta^\mu_{~\nu} ~~, ~~ \mu , \nu = 0,\ldots, p~~~;~~~R^\mu_{~\nu}  = -  \delta^\mu_{~\nu} ~~, ~~ \mu , \nu = p+1,\ldots,9~.
\end{equation}
In this way one obtains:
\begin{eqnarray}
\frac{1}{2} T_p \kappa_{10} \,\,A_{\nu}\frac{ \left(  R^{\nu {\bar{\nu}}} + \frac{3-p}{4} \eta^{\nu {\bar{\nu}}}   \right)}{(-t)}  {\bar{A}}_{{\bar{\nu}}} 
\label{inelampl}
\end{eqnarray}
where $2 \kappa_{10}^{2} = (2\pi)^7 g^2 (\alpha')^4$, A and ${\bar{A}}$ stand for one of the two amplitudes of the previous subsection and 
$t = - p_3^2 = -(p_1 +p_2)^2$ is the momentum transfer in the inelastic process. It is easy to check that 
\begin{eqnarray}
\frac{1}{2} T_p \kappa_{10} =  \frac{ \pi^{\frac{9-p}{2}} R_{p}^{7-p}}{\Gamma ( \frac{7-p}{2} )}
\label{TpkaR}
\end{eqnarray}
appearing in Eq. (\ref{disk}).
Let us consider the case where both the right and left three-point amplitudes  are as in Eq. 
(\ref{V001covz}). We get:
\begin{eqnarray}
&& \frac{1}{2} T_p \kappa_{10} \,\,
\epsilon^{IJ}_{\rho \alpha} \left\{ \frac{\alpha'}{2} \left[ \eta^{\mu \alpha} k_{1}^{\nu} +   \eta^{\nu \alpha} q^{\mu} - \eta^{\mu \nu} q^{\alpha}\right] q^{\rho} - \eta^{\mu \alpha} \eta^{\nu \rho} \right\} 
\left( R_{\nu {\bar{\nu}}} + \frac{3-p}{4} \eta_{\nu {\bar{\nu}}} \right) \nonumber \\
&&\epsilon^{ \bar{I} {\bar{J}} }_{{\bar{\rho}} {\bar{\alpha}}} \left\{ \frac{\alpha'}{2}\left[ \eta^{{\bar{\mu}} {\bar{\alpha}}} k_{1}^{{\bar{\nu}}} +   \eta^{{\bar{\nu}} {\bar{\alpha}}} q^{{\bar{\mu}}} - \eta^{{\bar{\mu}} {\bar{\nu}}} q^{{\bar{\alpha}}}\right] q^{{\bar{\rho}}} - \eta^{{\bar{\mu}} {\bar{\alpha}}} \eta^{{\bar{\nu}} {\bar{\rho}}} \right\}
\label{A1A1v}
\end{eqnarray}
The term $\frac{\alpha'}{2} k_1 R k_1= (-\alpha's)$ gives  a divergent term at high energy. Furthermore, we have to remember that in the case of a massive state the longitudinal polarization is also enhanced at high energy. Taking this into account  we get the following amplitude:
\begin{eqnarray}
\frac{1}{2} T_p \kappa_{10} \,\,
\epsilon^{IJ}_{\rho \alpha}  \epsilon^{ \bar{I} {\bar{J}} }_{{\bar{\rho}} {\bar{\alpha}}}
 (-\alpha' s)  \frac{\alpha'}{2}  \left[ \eta^{\mu \alpha}  \left(q^{\rho} - \frac{v^{\rho}}{\sqrt{\alpha'}}  \right) +  \frac{\alpha'}{2} q^{\rho} q^{\mu} 
  \frac{  v^{{\alpha}}}{\sqrt{\alpha'}}  \right]   \left[ \eta^{{\bar{\mu}} {\bar{\alpha}}}  \left(q^{{\bar{\rho}}} - \frac{v^{{\bar{\rho}}}}{\sqrt{\alpha'}}  \right) + \frac{\alpha'}{2} q^{{\bar{\rho}}} q^{{\bar{\mu}}} 
  \frac{  v^{\bar{\alpha}}}{\sqrt{\alpha'}}  \right] \nonumber \\
\label{ththh}
\end{eqnarray} 
If we use the two amplitudes  as those in Eq. (\ref{3psiam}), we get
\begin{eqnarray}
&&\frac{1}{2} T_p \kappa_{10} \,\,
\epsilon^{IJK}_{\rho \sigma \tau} \left[  q^{\rho} \left( \eta^{\nu \sigma} \eta^{\mu \tau} -
\eta^{\mu \sigma} \eta^{\nu \tau}  \right)  + q^{\sigma} \left( \eta^{\mu \rho} \eta^{\nu \tau} - 
 \eta^{\nu \rho} \eta^{\mu \tau}   \right) + q^{\tau} \left(\eta^{\nu \rho} \eta^{\mu \sigma} - 
 \eta^{\mu \rho} \eta^{\nu \sigma} \right) \right]  \nonumber \\
 && \times \left( R_{\nu {\bar{\nu}}} + \frac{3-p}{4} \eta_{  \nu  { \bar{\nu}}} \right)  \nonumber \\
 && \times \frac{\alpha'}{2}\epsilon^{{\bar{I}}{\bar{J}}{\bar{K}}}_{{\bar{\rho}} {\bar{\sigma}} {\bar{\tau}}} \left[  q^{{\bar{\rho}}} \left( \eta^{{\bar{\nu}} {\bar{\sigma}}} \eta^{{\bar{\mu}} {\bar{\tau}}} -
\eta^{{\bar{\mu}} {\bar{\sigma}}} \eta^{{\bar{\nu}} {\bar{\tau}}}  \right)  + q^{{\bar{\sigma}}} \left( \eta^{{\bar{\mu}} {\bar{\rho}}} \eta^{{\bar{\nu}} {\bar{\tau}}} - 
 \eta^{{\bar{\nu}} {\bar{\rho}}} \eta^{{\bar{\mu}} {\bar{\tau}}}   \right) + q^{{\bar{\tau}}} \left(\eta^{{\bar{\nu}} {\bar{\rho}}} \eta^{{\bar{\mu}} {\bar{\sigma}}} - 
 \eta^{{\bar{\mu}} {\bar{\rho}}} \eta^{{\bar{\nu}} {\bar{\sigma}}} \right) \right] \nonumber \\
\label{psipsi3}
\end{eqnarray}
Taking again into account the enhancement at high energy due to the longitudinal polarization
one gets:
\begin{eqnarray}
&&\frac{1}{2} T_p \kappa_{10} \,\,  \frac{(-\alpha's)}{2} \frac{\alpha'}{2}
\epsilon^{IJK}_{\rho \sigma \tau} \left[  q^{\rho} \left( \eta^{\mu \sigma} v^{\tau} -
\eta^{\mu \tau} v^{\sigma}  \right)  + q^{\sigma} \left( \eta^{\mu \rho} v^{\tau} - 
 v^{\rho} \eta^{\mu \tau}   \right) + q^{\tau} \left(\eta^{\mu \rho} v^{\sigma} - 
 v^{\rho} \eta^{\mu \sigma} \right) \right] \nonumber \\
&& \epsilon^{{\bar{I}}{\bar{J}}{\bar{K}}}_{{\bar{\rho}} {\bar{\sigma}} {\bar{\tau}}}  \left[  q^{{\bar{\rho}}} \left( \eta^{{\bar{\mu}} {\bar{\sigma}}} v^{{\bar{\tau}}} -
\eta^{{\bar{\mu}} {\bar{\tau}}} v^{{\bar{\sigma}}}  \right)  + q^{{\bar{\sigma}}} \left( \eta^{{\bar{\mu}} {\bar{\rho}}} v^{{\bar{\tau}}} - 
 v^{{\bar{\rho}}} \eta^{{\bar{\mu}} {\bar{\tau}}}   \right) + q^{{\bar{\tau}}} \left(\eta^{{\bar{\mu}} {\bar{\rho}}} v^{{\bar{\sigma}}} - 
 v^{{\bar{\rho}}} \eta^{{\bar{\mu}} {\bar{\sigma}}} \right) \right] \nonumber \\
\label{psi3xxx}
\end{eqnarray}
Using the kinematics of the Appendix one can write  the quantity in one of the two squared 
brackets in Eq. (\ref{ththh}) as follows:
\begin{eqnarray}
A^{IJ}_{k} &=& \epsilon^k_{\mu} \epsilon^{IJ}_{\rho \alpha}   \left[ \eta^{\mu \alpha}  \left(q^{\rho} - \frac{v^{\rho}}{\sqrt{\alpha'}}  \right) +  \frac{\alpha'}{2} q^{\rho} q^{\mu}   \frac{  v^{{\alpha}}}{\sqrt{\alpha'}}  \right]  \nonumber \\
&=&  \frac{1}{2} \left[ {\bar{p}}_{1}^{I} \delta^{kJ} + {\bar{p}}_{1}^{J} \delta^{kI} - \frac{1}{3}
{\bar{p}}_{1}^{k} \delta^{IJ}    \right] + \frac{1}{2} {\bar{p}}_{1}^{k}  \delta^{Iv} \delta^{Jv}
\label{AIJk}
\end{eqnarray}
where $k= 1 \dots 8; I,J= 1 \dots 8, v$. If we divide the 9-dim indices $I= (i, v)$ and $J=(j,v)$ in an 8-dim part and a part along $v$, from the previous expression we get:
\begin{eqnarray}
A^{ij}_{k} &=& \frac{1}{2} \left[ {\bar{p}}_{1}^{i} \delta^{kj} + {\bar{p}}_{1}^{j} \delta^{ki} \right]~~~;~~~ i \neq j \nonumber \\
A^{ii}_{k} &=& {\bar{p}}_{1}^{k} \left( \delta^{ik} - \frac{1}{6} \right) ~~;~~~i=1 \dots 8
\nonumber \\
A^{vv}_{k} &=&  - \sum_{i=1}^{8} A^{ii}_{k} =  \frac{1}{3}{\bar{p}}_{1}^{k}   \nonumber \\
A^{iv} &=& A^{vi} = 0
\label{AIJkijv}
\end{eqnarray}
Performing the same analysis with the antisymmetric amplitude in Eq. (\ref{psi3xxx}), we get:
\begin{eqnarray}
A^{IJH}_{k} =  \frac{1}{2} \epsilon_{\mu}^{k} \epsilon^{IJH}_{\rho \sigma \tau} \left[  {\bar{p}}_{1}^{\rho} \left( \eta^{\mu \sigma} {v^{\tau}} -
\eta^{\mu \tau}  {v^{\sigma}}   \right)  - {\bar{p}}_{1}^{\sigma} \left( \eta^{\mu \rho} 
{v^{\tau}} - { v^{\rho} }  \eta^{\mu \tau}   \right) + {\bar{p}}_{1}^{\tau} \left(\eta^{\mu \rho} 
 { v^{\sigma} } -  {v^{\rho}} \eta^{\mu \sigma} \right) \right] 
\label{AIJkv}
\end{eqnarray}
that implies
\begin{eqnarray}
A^{ijh}_{k} &=&  A^{ivv} = 0 \nonumber \\
A^{ijv}  &=& \frac{1}{2} \left( {\bar{p}}_1^{i} \delta^{K j} - {\bar{p}}_1^{j} \delta^{K i}    \right)
\label{Aijkv}
\end{eqnarray}
Remembering the connection between covariant and light-cone states, from the previous expressions we see that the scattering of a graviton on a $Dp$-brane will produce only closed string  states with left or right movers of the type $A_{-1;j} B_{-\frac{1}{2};k } |0 \rangle $ in the RNS case corresponding to the states  $A_{-1;j} |k \rangle$ in the GS case, while the states with left or right movers of the type  $B_{-\frac{1}{2};j} |0  \rangle $ and to $B_{-\frac{1}{2};i} B_{-\frac{1}{2};j} B_{-\frac{1}{2};k} |0\rangle$, corresponding to   $Q_{-1;i} |j \rangle$  in the GS case, are not produced at high energy.  This is in agreement with what one gets from the eikonal operator interpreting 
the bosonic oscillators as the string bosonic oscillators in the light-cone gauge. In the next section we will derive the eikonal operator directly from string theory without needing to go through the scattering amplitude and require unitarity as it was done in Sect. \ref{eikonal,I}.

\section{The eikonal operator  II}
\label{eikonal,II}

In this section we sketch the construction of the eikonal operator that was done in Ref.~\cite{ddrv2}. 
The first ingredient  is the GS three-string vertex given by:
\begin{equation}
\label{eq:Vs}
  |V_{GS} \rangle = \left(P_i-\alpha_1 \alpha_2 \alpha_3 \frac{n}{\alpha_q} N^q_n A^q_{-n, i} \right) V_b V_f | V_i \rangle  |V_0 \rangle ~,
\end{equation}
where
 \begin{eqnarray}
  V_b &=& \exp\left(\frac 12 A^p_{-n, i} N^{pq}_{mn} A^q_{-m, i} + P_i N^q_n A^q_{-n, i} \right)~,\nonumber \\
  V_f &=& \exp\left(\frac 12 Q^p_{-n, a} X^{pq}_{mn} Q^q_{-m, a} - S_a \frac{n}{\alpha_q} N^q_n Q^q_{-n, a} \right)~, \nonumber \\
  |V_i\rangle &=& \frac{1}{\alpha_1} |ijj\rangle + \frac{1}{\alpha_2} |jij\rangle  + \frac{1}{\alpha_3} |jji\rangle + \frac{\alpha_1-\alpha_2}{4 \alpha_3} |aai\rangle + \frac{\alpha_1-\alpha_3}{4 \alpha_2} |aia\rangle  \nonumber \\
&+& \frac{\alpha_2-\alpha_3}{4 \alpha_1} |iaa\rangle + \frac{1}{4} \gamma^{ij}_{ab} \left(|baj\rangle + |bja\rangle + |jba\rangle \right)~.
 \label{eq:VbVf}
\end{eqnarray}
To insure momentum conservation we have included in the vertex    a part with  
the bosonic zero modes given by:
\begin{eqnarray}
|V_0 \rangle = \int d^{10} x \,\, |x\rangle_1 \,\, |x\rangle_2 \,\, |x\rangle_3 = (2\pi)^{10} \delta^{(10)} ({\hat{p_1}} + {\hat{p_2}} + {\hat{p_3}}) |x=0\rangle_1 \,\,  |x=0\rangle_2 \,\,  |x=0\rangle_3 
\label{V0}
\end{eqnarray}
where the state $|x \rangle$ is an eigenstate of the position operator: ${\hat{q}} |x\rangle = x |x\rangle$.
The operators $P_i$ and $S_a$ stand for the following combinations of
the bosonic and fermionic zero-modes
\begin{equation}
  \label{eq:Pi}
  P_i \equiv  \left(\alpha_r \bar{p}^{(r+1)}_{iL} - \alpha_{r+1} \bar{p}^{(r)}_{iL}\right)~, ~~~~
  S_a \equiv \alpha_r Q^{(r+1)}_{0 a} - \alpha_{r+1} Q^{(r)}_{0 a}~.  
\end{equation}
which, with the cyclic identification between $r=4$ and $r=1$, are
independent of the choice of $r=1,2,3$. Finally, the `Neumann'
coefficients encoding the actual value of the various couplings are
\begin{gather}
\label{NMX}
N^{rs}_{nm} =  - \frac{nm \alpha_1 \alpha_2 \alpha_3}{n \alpha_s + m
  \alpha_r} N^{r}_{n} N_{m}^{s} ~;~~~
  X^{rs}_{nm} = \frac{n \alpha_s - m\alpha_r}{2\alpha_r \alpha_s}
  N^{rs}_{nm}~, \\ \label{NMX2}
N_{n}^{r} = - \frac{1}{n \alpha_{r+1}}\left(  \begin{array}{c} - n \frac{\alpha_{r+1}}{\alpha_r} \\ n \end{array} \right) = \frac{1}{\alpha_r n!} \frac{ \Gamma  \left(   - n 
\frac{\alpha_{r+1}}{\alpha_r} \right)  }{ \Gamma  \left(   - n 
\frac{\alpha_{r+1}}{\alpha_r} +1 -n \right) }~.
\end{gather}
Remember that the light-cone three-string vertex  depends on a light-like vector $k$ that in general can be chosen as we want. It turns out, however, that, if we choose it to be along the direction of the two energetic strings, at high energy the vertex gets enormously  simplified. Since we have chosen the momentum of incoming graviton and massive state as  in Eqs. (\ref{p2v}) and (\ref{p1xxx}), 
this means that we have to choose   $k= \frac{1}{\sqrt{2}}(-1, 0_p;0_{8-p}, 1)$. Momentum conservation implies that the momentum of the third string is given by $p_2 = (0,0_p; -{\bar{p}}_1, -q_9)$~\footnote{Notice that the state labelled here by $r=3$ has momentum $p_2$ in (\ref{p2v}).}.
Proceeding in this way, at high energy,  we get the following GS vertex:
 \begin{equation}
  \label{eq:Vss}
  |V_{GS}\rangle \sim \frac{P_i}{\alpha_2}
   \exp\left\{-\sqrt{\frac{\alpha'}{2}} \frac{{\bar{p}}_{1 \ell}}{n} \Big({A}^3_{-n \ell}+(-1)^n {A}^1_{-n \ell}\Big)\right\} \left[
    |jij\rangle + \frac{\alpha_1-\alpha_3}{4} |aia\rangle \right] ~. 
\end{equation}
The second ingredient is the  boundary state in the light-cone gauge that was constructed in Ref.~\cite{Green:1996um}.  We use a slightly modified version of it where we impose Neumann (Dirichlet) boundary conditions along the longitudinal (transverse) directions to the world volume of the $Dp$-branes. 
It is given by:
 \begin{eqnarray}
| B, \eta, y \rangle \sim    \exp \left\{-
\sum_{n=1}^{\infty}  \left[  \frac{1}{n} {\alpha_{-n}^{i} D_{ij}  {\tilde{\alpha}}_{-n}^{j}}
+ i \eta S_{-n}^{a} M_{a \dot{b}} {\tilde{S}}^{\dot{b}}_{-n} \right]  \right\}| B_0, \eta,y \rangle
\label{bstateb}
\end{eqnarray}
 where $R$ is the reflection matrix given in Eq. (\ref{reflema}),
\begin{eqnarray}
| B_0, \eta,y \rangle=  \left(R_{ij} |i \rangle | {\tilde{j}}\rangle + 
i \eta M_{\dot{a} b} |\dot{a} \rangle | {\tilde{b}}\rangle\right)  \delta^{(9-p)} (\hat{q} - y) 
 |0_\alpha, p=0\rangle 
\label{B0Dx}
\end{eqnarray}
and
\begin{eqnarray}
M_{\dot{a} {b}} = i \left( \gamma^1 \gamma^2 \dots \gamma^{p+1} \right)_{\dot{a} b}~~;~~
M_{a \dot{b}}=  i \left( \gamma^1 \gamma^2 \dots \gamma^{p+1} \right)_{{a} \dot{b}}~~.
\label{MMD}
\end{eqnarray}
The third ingredient is the light-cone propagator:
\begin{eqnarray}
P = \frac{ \pi \alpha' }{2} 
\int_{0}^{\infty} dt \,\, {\rm e}^{-\pi t \left( \frac{\alpha'}{2} {\hat{p}}^{2}_{i} + N + {\tilde{N}}   \right)}
~~;~~i =1 \dots 8
\label{PPcc}
\end{eqnarray}
where $N$ and ${\tilde{N}}$ are the bosonic and fermionic number operators. 

Using the three previous ingredients, we compute the quantity:
\begin{equation}
  \label{eq:BPV}
  \frac{T_p}{2} ~ {}_2\langle B| P \left(\kappa_{10}  |V_{GS} \rangle
    |\widetilde{V_{GS}} \rangle \right) \sim \frac{R_{p}^{7-p}
    \pi^{\frac{9-p}{2}}}{\Gamma \left(\frac{7-p}{2} \right)}~
  {}_2\langle B_0| \frac{1}{-t} \left( |V_{GS}
  \rangle |\widetilde{V_{GS}} \rangle \right)~. 
\end{equation}
In particular, in the previous equation we limit ourselves only to the pole of the graviton, as we have done in the previous section. Then we can neglect all oscillators in the boundary state and in the propagator and we need  only to consider the contribution of the bosonic zero modes:
\begin{eqnarray}
{}_2 \langle p=0 | \,\,\delta^{9-p} ({\hat{q}})\,\,  \frac{1}{{\hat{p}}_i^2} \,\, | x\rangle_2 =
{}_2 \langle p=0 | \,\, \int \frac{d^{9-p} k}{(2\pi)^{9-p}} {\rm e}^{i k \cdot {\hat{q}}}
\,\,  \frac{1}{{\hat{p}}_i^2} \,\, | x\rangle_2 =  \int \frac{d^{9-p} k}{(2\pi)^{9-p}} \frac{{\rm e}^{i k \cdot x}}{k_i^2}
\label{boszermod}
\end{eqnarray}
Then, assuming that the strings $1$ and $3$ have momentum $p_1$ and $p_3$, we get $(\langle x | p\rangle = {\rm e}^{- ipx})$:
\begin{eqnarray}
\int d^{10}  x  \langle p_1 | x \rangle_1  \langle p_3 | x \rangle_3  \int \frac{d^{9-p} k}{(2\pi)^{9-p}} \frac{{\rm e}^{i k \cdot x}}{k_i^2} = (2\pi)^{p+1} \delta^{(p+1)} (p_1 + p_3) \frac{1}{(-t)}
\label{mome}
\end{eqnarray}
where $t = - (p_1+p_3)^2$ is the momentum transfer.  Using the following equation~\cite{ddrv2}:
\begin{equation}
  \label{eq:PDP}
  \frac{2}{\alpha'} \frac{P_h R_{hk} P_k}{\alpha_{2}^{2} (-t)} =
  \frac{\alpha_3^2}{\alpha_2^2} \frac{({\bar{p}}_1)^2}{ t} = -
  \frac{\alpha_3^2}{ \alpha_2^2}  \left( 1 + \frac{q_{9}^{2}}{t}\right)
  \sim - \frac{4E^2}{q_{9}^{2}} - \frac{4E^2}{t}~,
\end{equation}
and neglecting the term without the pole at $t=0$ we arrive at
\begin{eqnarray}
  |W\rangle & \sim & 
  \frac{R_{p}^{7-p} \pi^{\frac{9-p}{2}}}{\Gamma
    \left(\frac{7-p}{2} \right)}
  \frac{4 E^2}{-t} \exp\left\{-\sqrt{\frac{\alpha'}{2}} \frac{{\bar{p}}_{1 \ell}}{n}
    ({A}^3_{-n \ell}+(-1)^n  {A}^1_{-n \ell})\right\} 
  \left[|j\rangle_1 |j\rangle_3 + \frac{\alpha_1}{2} |a\rangle_1
    |a\rangle_3 \right]
  \nonumber \\ \label{eq:zmVD}  && 
   \times ~ \exp\left\{-\sqrt{\frac{\alpha'}{2}} \frac{{\bar{p}}_{1 \ell}}{n}
    (\tilde{A}^3_{-n \ell}+(-1)^n \tilde{A}^1_{-n \ell})\right\} 
  \left[|\tilde{j}\rangle_1 |\tilde{j}\rangle_3 + \frac{\alpha_1}{2}
    |\tilde{a} \rangle_1 |\tilde{a}\rangle_3 \right]~.
\end{eqnarray}
Following Ref.~\cite{ddrv2} we can finally write it in a single Hilbert space getting:
\begin{eqnarray}
 W & \sim & 
  \frac{R_{p}^{7-p} \pi^{\frac{9-p}{2}}}{\Gamma
    \left(\frac{7-p}{2} \right)}
  \frac{4 E^2}{-t} : \exp\left\{-\sqrt{\frac{\alpha'}{2}} \frac{{\bar{p}}_{1 \ell}}{n}
    ({A}_{-n \ell}-  {A}_{n \ell})\right\}: 
  \nonumber \\ \label{eq:zmVD1}  && 
   \times ~ : \exp\left\{-\sqrt{\frac{\alpha'}{2}} \frac{{\bar{p}}_{1 \ell}}{n}
    (\tilde{A}_{-n \ell}-  \tilde{A}_{n \ell})\right\} : .
\end{eqnarray}
Introducing an auxiliary string coordinate (without zero modes):
\begin{equation}
  \label{eq:hatX}
  \hat{X}^i(\sigma) = \ii \sqrt{\frac{\alpha'}{2}} \sum_{n\not=0} \left(\frac{A_{ni}}{n} \ex{\ii n \sigma} + \frac{\tilde{A}_{ni}} {n} \ex{-\ii n \sigma}\right) ~.
\end{equation}
we can write~\eqref{eq:zmVD1} in an operator form as follows
\begin{equation}
  \label{eq:zmVDop}
  W(\bar{p}_1) = \int_{0}^{2 \pi} \frac{d\sigma}{2\pi} :\ex{\ii \bar{p}_1
    \hat{X}(\sigma)}:  \left(\frac{R_{p}^{7-p} \pi^{\frac{9-p}{2}}}{\Gamma
    \left(\frac{7-p}{2} \right)}\frac{4E^2}{-t}\right)~,
\end{equation}
that provides the same amplitude as in Eq. (\ref{eq:zmVD1}) when we saturate them with physical states satisfying the level matching condition. This operator is identical to the eikonal operator in Eq. (\ref{ep}) if we take the limit $\alpha' \rightarrow 0$ in the amplitude ${\cal{A}}_1$  given in Eq.
(\ref{T1}).  The $\alpha'$ corrections are recovered if one does not  include  just the contribution of the graviton as we have done above, but add also the contribution of the other string states.

In conclusion, we have provided two independent derivations of the eikonal operator. The one in this  section shows that the bosonic oscillators are the bosonic oscillators of superstring theory in a suitably chosen   light-cone gauge. This means that when we sandwich  the eikonal operator between two  arbitrary string states, we obtain  the production amplitude of one of them from the scattering of the other  on a $Dp$-brane at high energy and small transverse momentum.

\begin{appendix}

\section{Kinematics}
\label{kinematics}

The scattering amplitude for the production of a massive string  with momentum $p_2$ from the scattering of a graviton with momentum $p_1$ on a $Dp$-brane is described by the two 
(Mandelstam like) variables:
\begin{equation}
  \label{eq:s-t}
  t= - q^2 = -(p_1 + p_2)^2~,~~~
  s= -\frac{1}{4} (p_1 + R p_1)^2= -\frac{1}{4} (p_2 + R p_2)^2\equiv E^2~,
\end{equation}
where in the second equation we used the momentum conservation along
the Neumann directions and $E > 0$ will denote, hereafter, the common energy of the  incoming and outgoing closed strings. It is convenient to choose the massive string to move along the $9$-th space direction:
\begin{equation}
 p_{2}^{\mu} = \left( -E , 0_p ; 0_{8-p} , - \sqrt{E^2 -M^2} \right) ~,
\label{p2v}
\end{equation}
where the first $p+1$ directions are parallel to the (Neumann
directions of the) D$p$-branes and the entries after the semicolon are
along the Dirichlet directions. The most direct way to describe the
physical polarization of massive particles is to introduce 9 vectors
perpendicular to their momentum. For instance, in the case of the
outgoing state~\eqref{p2v} we have the unit vectors $\hat{w}^i$
\begin{equation}
  \label{eq:versorw}
  \hat{w}_1 = \left(0,1, 0_{p-1}; 0_{8-p}, 0 \right)~,
  \ldots,~~\hat{w}_8 =\left(0, 0_{p}; 0_{7-p} 1, 0 \right) 
\end{equation}
and, as the ninth one, $v^{\mu}$ corresponding to the longitudinal
polarization: 
\begin{equation}
  \label{eq:long}
 v^{\mu}_2 =  \left( \frac{\sqrt{E^2 -M^2}}{M} , 0_p ; 0_{8-p} ,
   \frac{E}{M}  \right) ~.   
\end{equation}
The possible momenta of the ingoing massless string  take the following form
\begin{gather}
  p_1^\mu = \left( E,0_p;{\bar{p}}_1, \sqrt{E^2-M^2} + q_9 \right)~,
\label{p1xxx}
\\
q^{\mu=9} =\frac{t+M^2}{2 \sqrt{E^2-M^2}}~,~~~~
({\bar{p}}_1)^{2} + (q^{\mu=9})^2 = -t \equiv (p_1+p_2)^2~,
\label{pbarq}
\end{gather}
where ${\bar{p}}_1$ is a $(8-p)$-dim vector orthogonal to the direction of motion of the massive string.  It is convenient to choose the eight polarizations of the massless string as follows:
\begin{equation}
\epsilon^{\mu}_{k} = \left(\frac{{\bar{p}}_{1}^{k}}{E+\sqrt{E^2-M^2} +
    q^9},  \delta^{i}_{k} , -\frac{{\bar{p}}_{1}^{k}}{E+\sqrt{E^2-M^2} +
    q^9} \right) 
\label{emuK}
\end{equation}
It is easy to check that $\epsilon^{\mu}_{k} p_{1\mu} =0$ for any $k=1 \dots 8$. Using this we can compute
\begin{eqnarray}
\epsilon_k  q \equiv \epsilon_k (p_1+p_2) = \epsilon_k p_2 = {\bar{p}}_1^k
\label{eq:eetaq}
\end{eqnarray}
where we have kept only the leading term at high energy.

\end{appendix}

\end{document}